\documentclass[12pt
  ]
{article}
\usepackage[swedish,english]{babel} % Use English and Swedish languages. English default. English hyphenation.
\usepackage[latin1]{inputenc} % Unix users should include this package instead of ansinew for correct handling of diacritics.
\usepackage{a4wide}

\usepackage{amsmath}
\usepackage{latexsym}
\usepackage{amssymb}

\usepackage{graphicx}
\graphicspath{{fig/}}
\usepackage[font=small,labelfont={small,bf},labelsep=period]{caption}

\newcommand{\ie}{{\em i.e.}}

\newcommand{\GeV}{\;\mathrm{GeV}}
\newcommand{\MeV}{\;\mathrm{MeV}}
\newcommand{\nutev}{NuTeV}

\DeclareCaptionLabelFormat{mine}{\small {\bf FIGURE #2}}
\newenvironment{rightcaption}{
\small
\setlength{\parindent}{0pt}
\addtocounter{figure}{1}
{\bf FIGURE \thefigure.}\hspace*{2pt}
}

\begin{document}

%%%%%%%%%%%%%%%%%%%%%%%%%%%%%%%%%%%%%%%%%%%%
%% FRONTMATTER
%%%%%%%%%%%%%%%%%%%%%%%%%%%%%%%%%%%%%%%%%%%%
\hfill TSL/ISV-2005-0294

\hfill August 2005

\vspace*{-0.5em}
\begin{center}
{\Large\bf Quark Asymmetries and Intrinsic Charm in Nucleons}%
\footnote{This is an extended version of the talk ``Quark Asymmetries in
Nucleons'', given at the XIII International Workshop on Deep Inelastic
Scattering, Madison, USA, April 27-May 1, 2005}\\[1em]
{\large Johan Alwall}\\[0.5em]
{\it High Energy Physics, Uppsala University, Box 535, S-75121 Uppsala, Sweden}
\end{center}

\begin{quotation}
\noindent{\bf Abstract:} We have developed a physical model for the
non-perturbative $x$-shape of parton density functions in the proton,
based on Gaussian fluctuations in momenta, and quantum fluctuations of
the proton into meson-baryon pairs. The model describes the proton
structure function and gives a natural explanation of observed quark
asymmetries, such as the difference between the anti-up and anti-down
sea quark distributions and between the up and down valence
distributions. We find an asymmetry in the momentum distribution of
strange and anti-strange quarks in the nucleon, large enough to reduce
the NuTeV anomaly to a level which does not give a significant
indication of physics beyond the standard model. We also consider
charmed fluctuations, and show that they can explain the excess at
large $x$ in the EMC $F_2^c$ data.\\[0.5em]
{\bf Keywords:} quark asymmetries, parton density distributions,
s-sbar asymmetry, NuTeV anomaly, intrinsic charm\\[0.5em]
{\bf PACS:} 12.39.Ki,11.30.Hv,12.40.Vv,13.15.+g,13.60.Hb
\end{quotation}
\vspace*{-0.5em}
%%%%%%%%%%%%%%%%%%%%%%%%%%%%%%%%%%%%%%%%%%%%
%% MAINMATTER
%%%%%%%%%%%%%%%%%%%%%%%%%%%%%%%%%%%%%%%%%%%%

%%%%%%%%%%%%%%%%%%%%%%
%\section{Introduction}
%%%%%%%%%%%%%%%%%%%%%%

The low-scale parton density functions give a description of the
hadron at a non-perturbative level. The conventional approach to these
functions is to make parameterizations using some more or less arbitrary
functional forms, based on data from deep inelastic scattering and
hadron collision experiments. Another approach, however, is to start
from some ideas of the behavior of partons in the non-perturbative
hadron, and build a model based on that behavior. The advantage with
this approach is that the successes and failures of such a model
allows us to get insight into the non-perturbative QCD dynamics.  The
model presented here, and described in detail in
\cite{Alwall:2005xd,Alwall:2004rd}, describes the $F_2$ structure
function of the proton, as well as sea quark asymmetries of the
nucleon. Most noteworthy, our model predicts an asymmetry between the
momentum distributions of strange and anti-strange quarks in the
nucleon of the same order as the newly reported results from NuTeV
\cite{Mason}. The model also suggests an intrinsic charm component in
the proton.

%%%%%%%%%%%%%%%%%%%%
%\section{The model}
%%%%%%%%%%%%%%%%%%%%

\begin{figure}
\includegraphics*[width=0.27\columnwidth,trim=0 70 0 0]{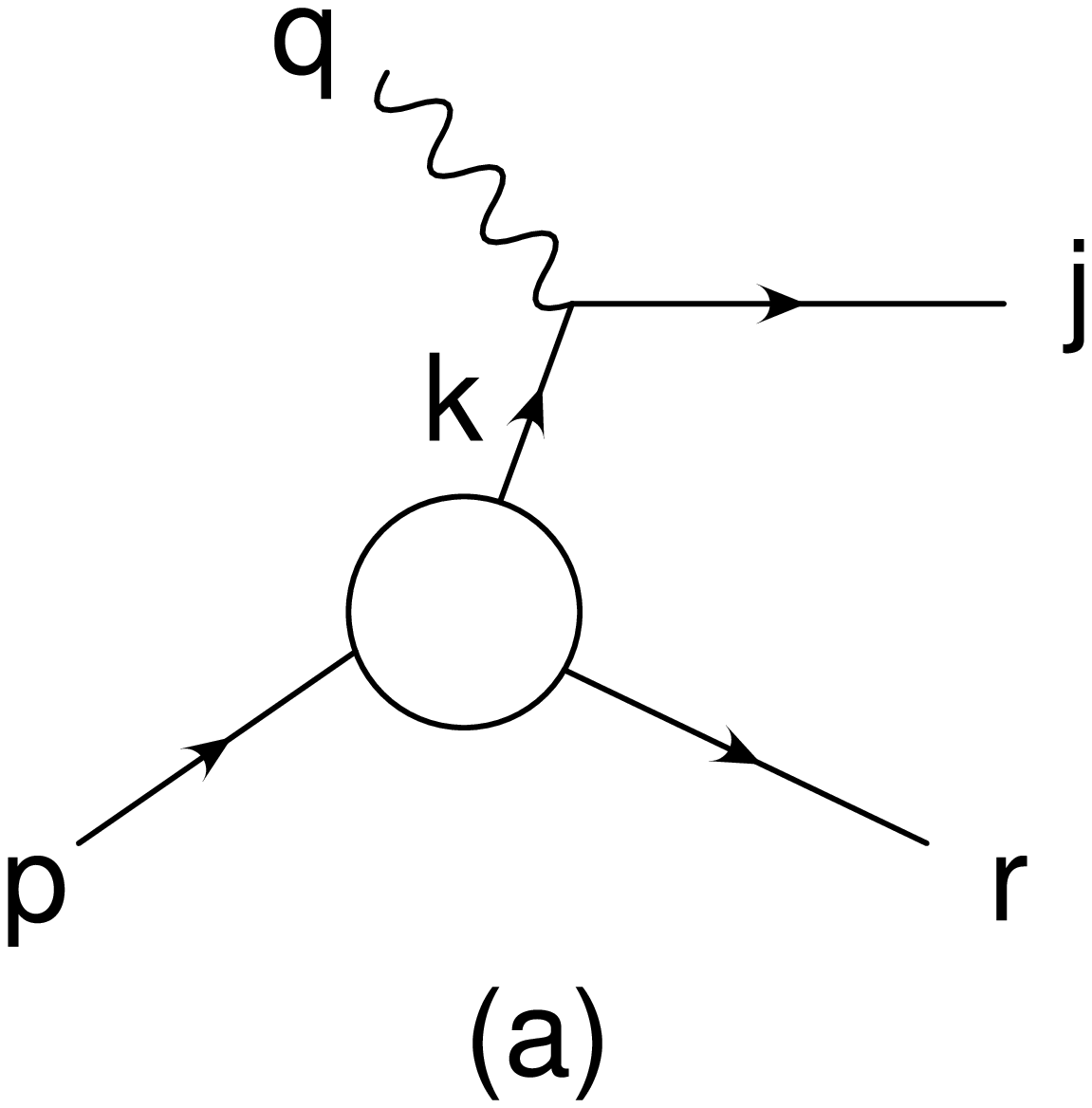}
\includegraphics*[width=0.27\columnwidth,trim=0 70 0 0]{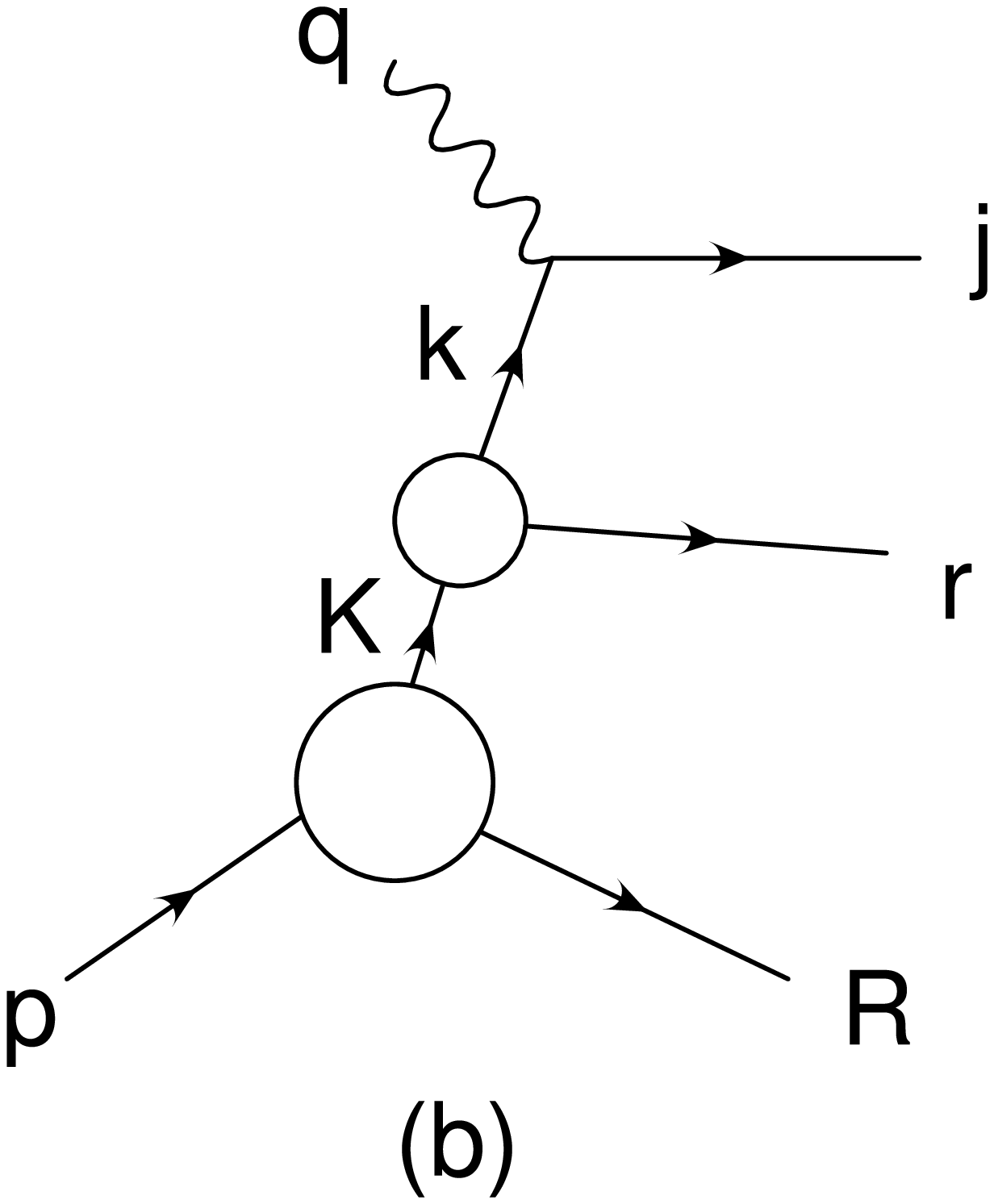}$\quad$
\includegraphics*[width=0.42\columnwidth,trim=0 10 0 0]{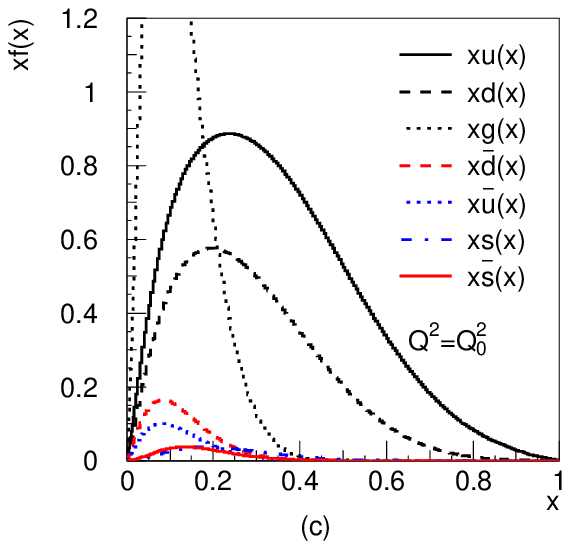}
\caption[FIGURE \thefigure.]{\label{fig:fluct} Illustration of the processes probing (a) a
valence parton in the proton and (b) a sea parton in a hadronic
fluctuation (letters are four-momenta). (c) shows the resulting parton
distributions at the starting scale $Q_0$.}
\end{figure}

This work extends the model previously presented in
\cite{Edin:1998dz}. The model gives the four-momentum $k$ of a single
probed valence parton (see Fig.~\ref{fig:fluct}a for definitions of
momenta) by assuming that, in the nucleon rest frame, the shape of the
momentum distribution for a parton of type $i$ and mass $m_i$ can be
taken as a Gaussian

\begin{equation}
f_i(k) = N(\sigma_i,m_i) \exp\left\{-\left[(k_0-m_i)^2+
k_x^2+k_y^2+k_z^2\right]/2\sigma_i^2\right\}
\end{equation}
which may be motivated as a result of the many interactions binding
the parton in the nucleon. The width of the distribution should be of
order hundred MeV from the Heisenberg uncertainty relation applied to
the nucleon size, \ie\ $\sigma_i=1/d_N$. The momentum fraction $x$ of
the parton is then defined as the light-cone fraction $x=k_+/p_+$. We
impose constraints on the final-state momenta in order to obtain a
kinematically allowed final state, which also ensures that $0<x<1$ and
$f_i(x)\to0$ for $x\to 1$. Using a Monte Carlo method these parton
distributions are integrated numerically without approximations.

To describe the dynamics of the sea partons, we note that the
appropriate basis for the non-perturbative dynamics of the bound state
nucleon should be hadronic. Therefore we consider hadronic
fluctuations, for the proton
\begin{equation} \label{eq:hadronfluctuation}
|p\rangle = \alpha_0|p_0\rangle + \alpha_{p\pi^0}|p\pi^0\rangle +
\alpha_{n\pi^+}|n\pi^+\rangle + \ldots + \alpha_{\Lambda K}|\Lambda
K^+\rangle + \ldots
\end{equation}
Probing a parton $i$ in a hadron $H$ of a baryon-meson fluctuation
$|BM\rangle$ (see Fig.~\ref{fig:fluct}b) gives a sea parton with
light-cone fraction $x=x_H\, x_i$ of the target proton. The momentum
of the probed hadron is given by a similar Gaussian, but with a
separate width parameter $\sigma_H$. Also here, kinematic constraints
ensure that we get a physically allowed final state. The procedure
gives $x_H\sim M_H/(M_B+M_M)$, \ie\ the heavier baryon gets a harder
spectrum than the lighter meson. The normalization of the sea
distributions is given by the normalization coefficients $\alpha_{BM}^2$ of
Eq.~\eqref{eq:hadronfluctuation}. These cannot be calculated from
first principles in QCD and are therefore taken as free parameters to
be fitted using experimental data.

The resulting valence and sea parton $x$-distributions apply at a low
scale $Q_0^2$, and the distributions at higher $Q^2$ are obtained
using perturbative QCD evolution at next-to-leading order.

The model has in total four shape parameters and three normalization
parameters, plus the starting scale, to determine the parton densities
$u$, $d$, $g$, $\bar{u}$, $\bar{d}$, $s$, $\bar{s}$. These are (with values
resulting from fits to experimental data as described below):
\begin{equation}
\label{eq:params}
\begin{array}{c}
\sigma_u=230\MeV \quad \sigma_d=170\MeV \quad 
\sigma_g=77\MeV \quad \sigma_H=100\MeV\\
\alpha_\mathrm{p\pi^0}^2=0.45 \quad \alpha_{n\pi^+}^2=0.14 \quad 
\alpha_{\Lambda K}^2=0.05 \quad Q_0=0.75\GeV
\end{array}
\end{equation}
The resulting parton densities are shown in Fig.~\ref{fig:fluct}(c).

%%%%%%%%%%%%%%%%%%%%%%%%%%%%%%%%%%%%%%%%%%%%%%%%%%
%\section{Comparison to data and parameter fitting}
%%%%%%%%%%%%%%%%%%%%%%%%%%%%%%%%%%%%%%%%%%%%%%%%%%

\begin{figure}
\center{\includegraphics*[width=0.8\columnwidth,trim=0 10 0 0]{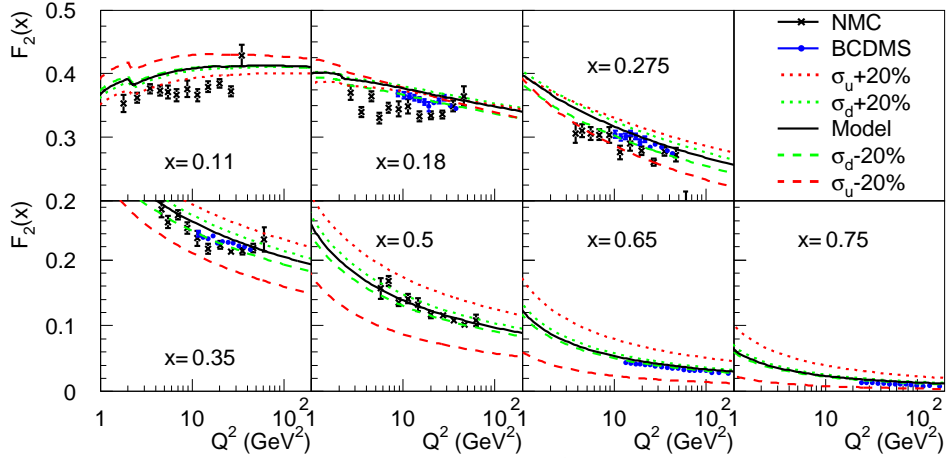}}
\caption{\label{fig:xbinned} The proton structure function
$F_2(x,Q^2)$ for large $x$ values; NMC and BCDMS data \cite{NMC,BCDMS}
compared to our model, also showing the results of $\pm 20\%$
variations of the width parameters $\sigma_u$ and $\sigma_d$ for the
$u$ and $d$ valence distributions.}
\end{figure}

\begin{figure}[b]
\begin{center}
\begin{minipage}{0.5\columnwidth}
\includegraphics*[width=0.8\columnwidth,trim=0 20 0 0]{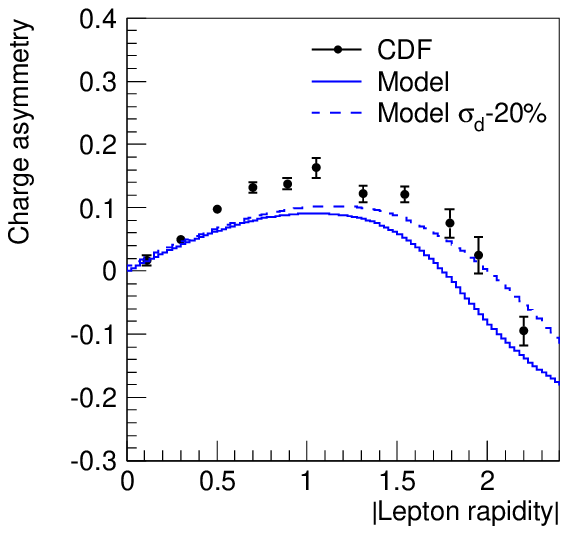}
\end{minipage}
\begin{minipage}{0.4\columnwidth}
\begin{rightcaption}
The charge asymmetry for
leptons from $W^\pm$-decays in $p\bar p$ collisions at the Tevatron
\cite{w-asym} compared to our model, with best-fit parameters and a
$20\%$ reduced width of the valence $d$ quark distribution.
\end{rightcaption}
\end{minipage}
\end{center}
\end{figure}

In order to fix the values of the model parameters, we make a global
fit using several experimental data sets: Fixed-target $F_2$ data to
fix large-$x$ (valence) distributions (Fig.~\ref{fig:xbinned}); HERA
$F_2$ data for the gluon distribution width and the starting scale
$Q_0$; $\bar d / \bar u$-asymmetry data for the normalizations of
the $|p\pi^0\rangle$ and $|n\pi^+\rangle$ fluctuations (see Fig.~4);
and strange sea data to fix the normalization of fluctuations
including strange quarks (see Fig.~5a). We have also compared with
$W^\pm$ charge asymmetry data as a cross-check on the ratio of
Gaussian widths for the $u$ and $d$ valence quark distributions
(Fig.~3). It is interesting to note that this simple model can
describe such a wealth of different data with just one or two
parameters per data set.

In our model, the shape difference between the valence $u$ and $d$
distributions in the proton, apparent from the $W^\pm$ charge
asymmetry data, is described as different Gaussian widths. This would
correspond to a larger effective volume in the proton for $d$ quarks
than for $u$ quarks, an effect which could conceivably be explained by
Pauli blocking of the $u$ quarks.

Since the proton can fluctuate to $\pi^0$ and $\pi^+$ by
$|p\pi^0\rangle$ and $|n\pi^+\rangle$, but to $\pi^-$ only by the
heavier $|\Delta^{++}\pi^-\rangle$, we get an excess of $\bar d$ over
$\bar u$ in the proton sea. Interestingly, the fit to data improves
when we use a larger effective pion mass of 400 MeV (see Fig.~4). This
might indicate that we have a surprisingly large coupling to heavier
$\rho$ mesons, or that one should use a more generic meson mass rather
than the very light pion.

\begin{figure}
\hspace*{-7mm}\begin{minipage}{0.7\columnwidth}
\includegraphics*[width=1\columnwidth,trim=0 0 0 0]{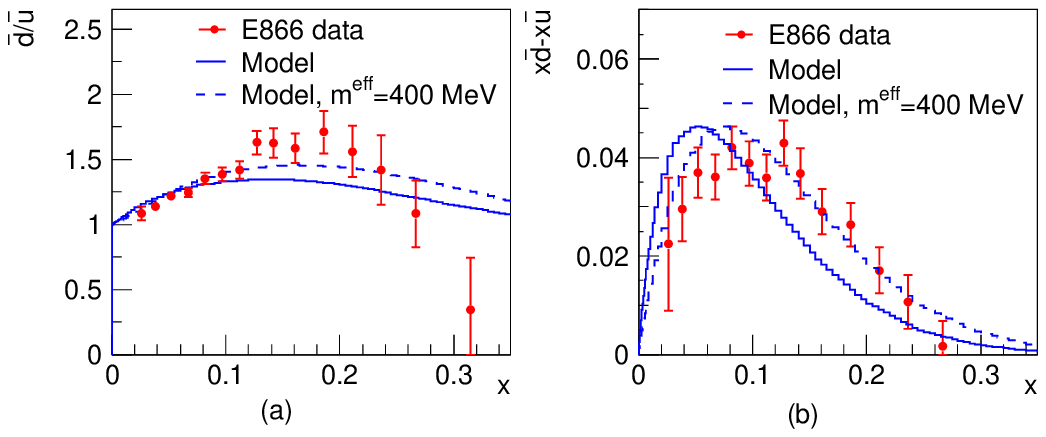}
\end{minipage}
\hspace*{5mm}\begin{minipage}{0.3\columnwidth}
\begin{rightcaption}
Comparison between our model and data from the E866/NuSea
collaboration \cite{dubar}: (a) $\bar u(x)/\bar d(x)$ (b)
$xd(x)-xu(x)$. The full line uses the physical pion mass, while the
dashed line uses an effective pions mass $m^\mathrm{eff} = 400\MeV$ as
discussed in the text.
\end{rightcaption}
\end{minipage}
\end{figure}

\begin{figure}[b]
\includegraphics*[width=0.6\columnwidth,trim=0 15 0 0]{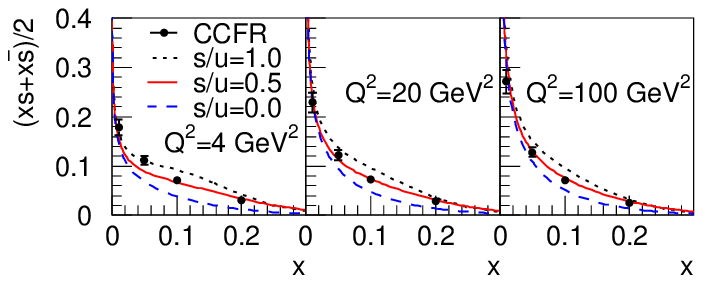}
\includegraphics*[width=0.4\columnwidth,trim=0 10 0 10]{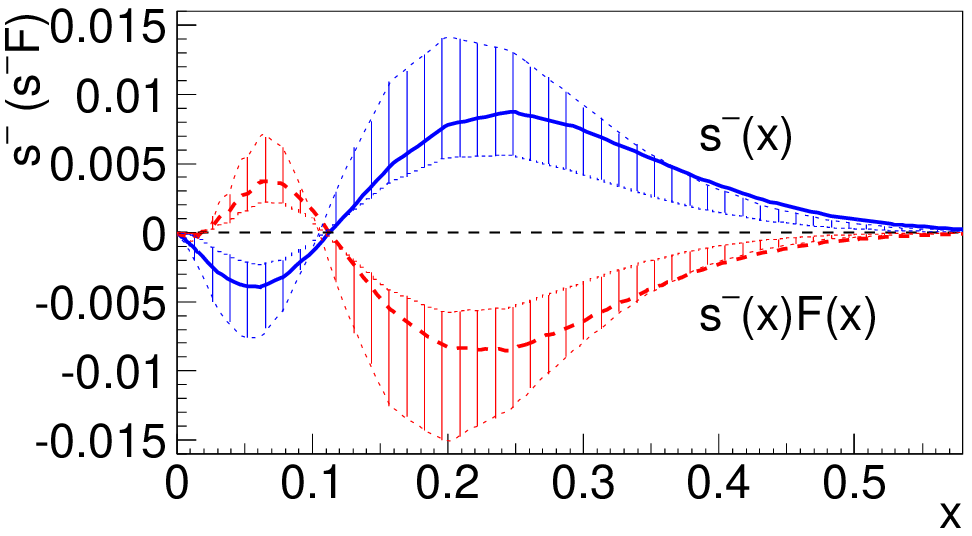}
\caption{\label{fig:sdata}(a) CCFR data \cite{ssbar} on the strange
sea distribution $(xs(x)+x\bar{s}(x))/2$ compared to our model based
on $|\Lambda K\rangle$ fluctuations with different normalizations.
(b) The strange sea asymmetry $s^-(x) = xs(x)-x\bar{s}(x)$ (at
$Q^2=20\GeV^2$) from the model and combined with the function $F(x)$
accounting for NuTeV's analysis giving $\Delta \sin^2\theta_W =
\int_0^1 dx\, s^-(x) F(x) = -0.0017$. The uncertainty bands correspond
to the uncertainties for $S^-$ and $\Delta \sin^2\theta_W$ quoted in the
text.}
\end{figure}

The lightest strange fluctuation is $|\Lambda K^+\rangle$. If we let
this implicitly include also heavier strange meson-baryon
fluctuations, we can fit the normalization $\alpha_{\Lambda K}^2$ to
strange sea data (see Fig.~5a). The result corresponds to
$\int_0^1(xs+x\bar s)dx/\int_0^1(x\bar u+x\bar d)dx \approx 0.5$, in
agreement with standard parton density parameterizations. We note that
this indicates a normalization $\propto 1/\Delta
M_{BM}=1/(M_B+M_M-M_p)$ rather than $\propto 1/\Delta M_{BM}^2$, as
expected from old-fashioned perturbation theory. The fluctuation
parameters are taken from the light sea results, $\sigma_H=100\MeV$
and $\sigma_q=\sigma_d^\mathrm{proton}=170\MeV$ as discussed in
\cite{Alwall:2005xd}.

Since the $s$ quark is in the heavier baryon $\Lambda$ and the $\bar
s$ quark is in the lighter meson $K^+$, we get a non-zero asymmetry
$S^- = \int_0^1 dx [xs(x)-x\bar s(x)]$ in the momentum distribution of
the strange sea, as seen in Fig.~5b and~6a. Depending on details of the
model, we get the range $ 0.0010\leq S^- \leq 0.0023$ for this
asymmetry. 

This is especially interesting in connection to the \nutev\
anomaly \cite{nutev}. \nutev\ found, based on the observable 
$R^- = \frac{\sigma(\nu_{\mu}N\to \nu_{\mu}X)-
            \sigma(\bar{\nu}_{\mu}N\to \bar{\nu}_{\mu}X)}
	   {\sigma(\nu_{\mu}N\to \mu^- X)-
            \sigma(\bar{\nu}_{\mu}N\to \mu^+X)}
   = g_L^2 - g_R^2 = \frac{1}{2} - \sin^2\theta_W$,
a $3\sigma$ deviation of $\sin^2\theta_W$ compared to the Standard
Model fit: $\sin^2\theta_W^\mathrm{NuTeV}=0.2277\pm 0.0016$ compared
to $\sin^2\theta_W^\mathrm{SM} = 0.2227\pm 0.0004$.  However, an
asymmetric strange sea would change their result, since $\nu$ only
have charged current interactions with $s$ and $\bar \nu$ with $\bar
s$. Using the folding function provided by \nutev\ to account for their
analysis, the $s$-$\bar s$ asymmetry from our model gives a shift
$-0.0024\leq\Delta\sin^2\theta_W=\int_0^1 dx\, s^-(x)
F(x)\leq-0.00097$, \ie\ the discrepancy with the Standard Model result
is reduced to between $1.6\sigma$ and $2.4\sigma$, leaving no strong
hint of physics beyond the Standard Model.

In our model, it is also natural to consider fluctuations involving
heavy quarks. Assuming that the hadronic fluctuation description is
still valid for the proton fluctuating into charmed baryon-meson
pairs, the lightest such fluctuations are $p\to\Lambda_C
\overline D$ and $p\to p\;J/\psi$. Taking these fluctuations into
account implies that there should be an intrinsic charm component in
the proton at intermediate $x\sim 0.4$. This component is quite
different from the purely perturbative charm component from gluon
splitting, which falls steeply with $x$ (see Fig.~7). The intrinsic
charm component should be present from the scale where the momentum
transfer can realize the charmed fluctuation.

\begin{figure}
\begin{minipage}{0.45\columnwidth}
\includegraphics[width=1\columnwidth]{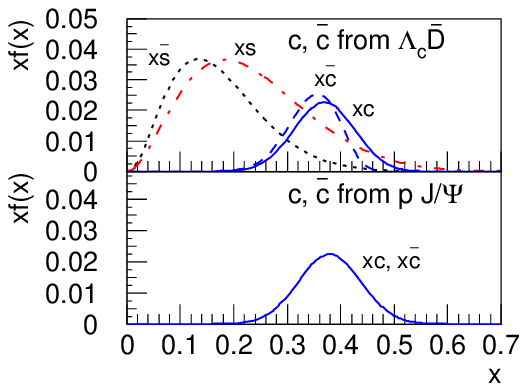}
\end{minipage}
\hspace*{5mm}\begin{minipage}{0.5\columnwidth}
\begin{rightcaption}
(a) Comparison between the strange and charm sea obtained from our
model with the inclusion of the $\Lambda_C\overline D$
fluctuation. The normalization is here taken to be $\propto
1/(M_B+M_M-M_p)$ as suggested by strange sea data. (b) The charm quark
distributions from the $p\;J/\psi$ fluctuation (the $c$ and $\bar c$
distributions are identical with this fluctuation). Note that the
distributions are very similar to those from the $\Lambda_C\overline D$
distribution, except for the small asymmetry between $c$ and $\bar c$.
\end{rightcaption}
\end{minipage}
\end{figure}

The resulting distributions of $c$ and $\bar c$ at the starting scale
are shown in Fig.~6. In Fig.~6a the distributions from the
$|\Lambda_C\overline D\rangle$ fluctuation are shown compared with the
$s$ and $\bar s$ distributions. The fluctuation parameters ($\sigma_H$
and $\sigma_q$) are taken to be the same as for the light quark
fluctuations, just as for the strange sea. However, the sensitivity of
the result on the precise values of these parameters is small. The
normalization is in Fig.~6 taken to be $\propto 1/\Delta M_{BM}$ (to
be discussed below), in order to easily compare the shapes of the
strange and charmed sea. In this case, there is an asymmetry between
the $c$ and $\bar c$ distributions, similar to that of the strange
sea, but much smaller due to the similarity in mass between the
$\Lambda_C$ ($2285\MeV$) and the $\overline D$ ($1865\MeV$). Fig.~6b
shows the distributions from the $|p\;J/\psi\rangle$ fluctuation. They
are very similar to that of the $|\Lambda_C\overline D\rangle$
distributions, except that the asymmetry is missing since both the $c$
and the $\bar c$ are here in the meson. From Fig.~6 it is clear that
an investigation of the intrinsic charm distributions from our model
is not much affected by the precise nature of the dominating
fluctuation mode, and in the following we will use only the
$|\Lambda_C\overline D\rangle$ fluctuation. Note that the shape of
our intrinsic charm component is somewhat different from that in the
intrinsic charm model by Brodsky et al.\ \cite{Brodsky:1980pb}, which
is based on partonic fluctuations $p\to uudc\bar c$ (see Fig.~7).

The normalization used in Fig.~6 corresponds to a relative importance
between different mass states proportional to $1/\Delta M_{BM}$, as
suggested by the strange sea fit to CCFR data. This normalization
would give an intrinsic charm component ($(c+\bar c)/2$ integrated number
density) of $0.9\%$. However, it might be more appropriate to use a
normalization $\propto 1/\Delta M_{BM}^2$ (as given by old-fashioned
perturbation theory) compared to the strange fluctuations,
corresponding to an intrinsic charm component of $0.18\%$.

\begin{figure}
\center{\includegraphics[width=0.8\columnwidth,trim=0 15 0 0]{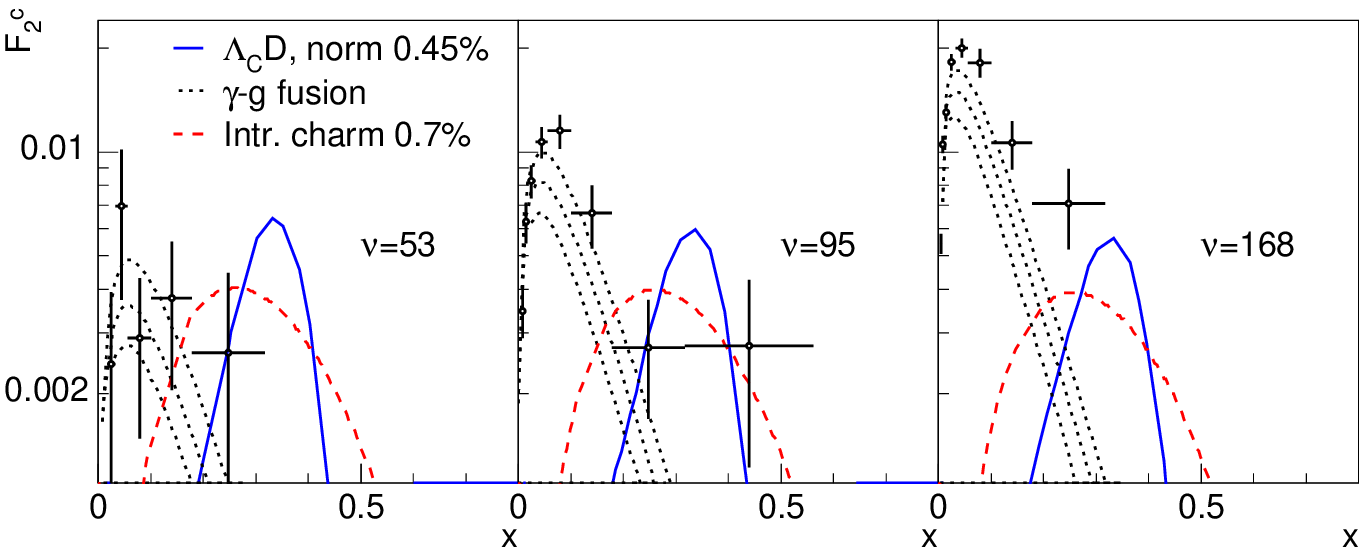}}
\caption{\label{fig:EMC} Data for the charm structure function $F_2^c$
from the EMC experiment \cite{Aubert:1982tt}, compared to our model
results with the best-fit normalization (solid curve). The
dotted curves show the perturbative QCD prediction for $\gamma g\to
c\bar c$ from \cite{Harris:1995jx}, with three different choices for
the renormalization and factorization scales. For comparison, the
intrinsic charm distribution of
\cite{Brodsky:1980pb} is also shown, with the 0.7\%
normalization allowed according to \cite{Harris:1995jx} (dashed curve).
}
\end{figure}

The only experimental data for the large-$x$ charmed structure
function $F_2^c = \frac49x(c+\bar c)$ (at leading order) comes from
the EMC experiment \cite{Aubert:1982tt}, which measured charmed hadron
production in muon-proton scattering. There, an intriguing excess was
found in the largest $x$ bins, compared to the perturbative
photon-gluon fusion expectation. A later analysis gave further
evidence that the excess cannot easily be attributed to standard
perturbative production channels \cite{Harris:1995jx}. Intrinsic charm
was immediately suggested as an explanation for the excess, but the
shape of the charm distribution in the original intrinsic charm model
is not optimal to explain the EMC excess. In Fig.~7, the EMC data is
shown in bins of $\nu=Q^2/2M_px$, together with the result from
our model, evolved in $Q^2$ using NLO QCD evolution \cite{Botje}. Here
we use the best-fit normalization, corresponding to $0.45\%$ intrinsic
charm.  This lies between the two normalizations discussed above,
which should not be surprising since the energy denominator only gives
an order-of-magnitude estimate. For comparison, we also show the
intrinsic charm distribution of \cite{Brodsky:1980pb} with the largest
normalization allowed by the EMC data according to
\cite{Harris:1995jx} ($0.7\%$ intrinsic charm), and the perturbative
photon-gluon fusion results from \cite{Harris:1995jx}.

As can be seen from Fig.~7, the shape of the intrinsic charm
distribution in our model seems to fit the data very well, giving an
enhancement at precisely the right values of $x$. Unfortunately, the
statistics of the EMC result is too small to allow any discrimination
between different models for intrinsic charm, and measurements of the
charm structure function at HERA (H1\cite{H1} and ZEUS\cite{ZEUS}) are
at too low values of $x$ to contribute to our understanding of
intrinsic charm. If a future experiment would measure the large-$x$
charm structure function with large statistics, it would be very
interesting to get a decisive verification of the presence of
intrinsic charm in the proton.

{\bf Acknowledgments: } I would like to thank the organizers for the
opportunity to talk at DIS'05, and Stan Brodsky and Gunnar Ingelman for
very interesting discussions.\vspace{-4mm}

%%%%%%%%%%%%%%%%%%%%%%%%%%%%%%%%%%%%%%%%%%%%%%%%
%% BACKMATTER
%%%%%%%%%%%%%%%%%%%%%%%%%%%%%%%%%%%%%%%%%%%%%%%%

%\begin{theacknowledgments}
%\end{theacknowledgments}



%\bibliographystyle{h-elsevier3}

%\bibliography{dis05_alwall}

\end{document}